\documentclass[aps,prd,12pt,preprint,superscriptaddress,preprintnumbers,nofootinbib,floatfix,longbibliography]{revtex4-1}

\usepackage[T1]{fontenc}
\usepackage[utf8]{inputenc}
\usepackage[english]{babel}
\usepackage{amsmath,amsthm,amssymb,amsfonts,mathrsfs,amsbsy,bm}
\usepackage{tensor}
\usepackage{slashed}
 \usepackage{mathrsfs,tikz} 
\usepackage{esint}
\usepackage[a4paper, margin=1.5cm]{geometry}
\usepackage{cancel}
\usepackage{placeins}
\usepackage{graphicx}
\usepackage{multirow}
\usepackage{array}
\usepackage{booktabs}
\usepackage{makecell}
\usepackage{xcolor}
\colorlet{BLUE}{blue}
\usepackage{tikz}
\usetikzlibrary{quotes,angles,arrows,decorations.markings,decorations.pathmorphing}
\usepackage{hyperref}
\hypersetup{colorlinks=true,breaklinks=true,citecolor=blue,linkcolor=[rgb]{0,0.5,0.9},urlcolor=blue}
\graphicspath{{figures/}}

\newcommand{\be}{\begin{equation}}
\newcommand{\ee}{\end{equation}}
\newcommand{\Be}{\begin{eqnarray}}
\newcommand{\Ee}{\end{eqnarray}}

\newcommand{\mincir}{\raise-3.truept\hbox{\rlap{\hbox{$\sim$}}\raise4.truept\hbox{$<$}\ }}
\newcommand{\magcir}{\raise-3.truept\hbox{\rlap{\hbox{$\sim$}}\raise4.truept\hbox{$>$}\ }}

\providecommand{\U}[1]{}

\newcommand{\ie}{\begin{equation}}
\newcommand{\fe}{\end{equation}}
\newcommand{\se}{\begin{eqnarray}}
\newcommand{\ff}{\end{eqnarray}}

\begin{document}
\emergencystretch=4em
\usetikzlibrary{arrows.meta,patterns}


\title{Optical appearance of a non--commutative black hole surrounded by an accretion disk}

\author{A. A. Ara\'{u}jo Filho}
\email{dilto@fisica.ufc.br}
\affiliation{Departamento de F\'isica, Universidade Federal da Para\'iba, Caixa Postal 5008, 58051--970, Jo\~ao Pessoa, Para\'iba, Brazil.}
\affiliation{Departamento de F\'isica, Universidade Federal de Campina Grande, Caixa Postal 10071, 58429--900 Campina Grande, Para\'iba, Brazil.}
\affiliation{Center for Theoretical Physics, Khazar University, 41 Mehseti Street, Baku, AZ-1096, Azerbaijan.}


\author{N.~Heidari}
\email{heidari.n@gmail.com }

\affiliation{Departamento de F\'isica, Universidade Federal de Campina Grande, Caixa Postal 10071, 58429--900 Campina Grande, Para\'iba, Brazil.}

\affiliation{Center for Theoretical Physics, Khazar University, 41 Mehseti Street, Baku, AZ-1096, Azerbaijan.}


\author{A. F. Santos}
\email{alesandroferreira@fisica.ufmt.br}
\affiliation{Programa de Pós-graduação em Física, Instituto de Física, Universidade Federal de Mato Grosso, Cuiabá, Brasil}


\author{S. Mamedov}
\email{sh.mamedov@khazar.org}
\affiliation{Institute for Physical Problems, Baku State University, Z.Khalilov 23, Baku, AZ 1148, Azerbaijan.}
\affiliation{Institute of Physics, Ministry of Science and Education, H.Javid 33, Baku, AZ 1143, Azerbaijan.}
\affiliation{Center for Theoretical Physics, Khazar University, 41 Mehseti Street, Baku, AZ-1096, Azerbaijan.}


\author{V. B. Bezerra}
\email{valdir@fisica.ufpb.br}
\affiliation{Departamento de F\'isica, Universidade Federal da Para\'iba, Caixa Postal 5008, 58051--970, Jo\~ao Pessoa, Para\'iba, Brazil.}

\author{Amilcar R. Queiroz}
\email{amilcarq@df.ufcg.edu.br}

\affiliation{Departamento de F\'isica, Universidade Federal de Campina Grande, Caixa Postal 10071, 58429--900 Campina Grande, Para\'iba, Brazil.}


\author{Ali \"Ovg\"un}
\email{ali.ovgun@emu.edu.tr}
\affiliation{
Physics Department, Eastern Mediterranean University, Famagusta, 99628 North Cyprus via Mersin 10, Turkiye.}

\date{\today}


\begin{abstract}

We investigate the optical appearance of a non--commutative Schwarzschild black hole surrounded by a transparent, geometrically thin emitting disk. A Lorentzian matter distribution supplies the effective exterior metric. We compute photon trajectories, disk transfer functions, and the observed bolometric intensity for static and circular emitters, taking into account gravitational redshift, orbital time dilation, and Doppler modulation. At fixed asymptotic mass, increasing $\theta=\Theta/M^2$ from zero to $0.03$ reduces the critical impact parameter by approximately $16.1\%$ and the innermost stable circular orbit radius by $23.7\%$. The narrow higher order features follow this contraction, while the broad direct peak shifts by less than $1\%$ for a fixed emission profile. With the same source normalization, the peak direct intensity and the signal integrated over the screen aperture increase by approximately $6.5\%$ and $4.3\%$, respectively. Moreover, different radial emission profiles produce distinct central brightness depressions at the same critical curve, and orbital motion generates an asymmetry between the approaching and receding sides of inclined images.

\end{abstract}

\maketitle



\section{Introduction}
\label{sec:introduction}

The images of M87* and Sgr~A* obtained by the Event Horizon Telescope have established the optical appearance of black holes as an observational probe of gravity in its strong regime~\cite{EHT2019,EHT2022SgrA}. Their bright emission and central depression reflect the spacetime geometry, the distribution of the emitting material, and its motion. In particular, unstable photon orbits determine the critical curve, while the visible boundary of the brightness depression also depends on the source~\cite{GrallaHolzWald2019,PerlickTsupko2022}. The frequency shift between the disk and the observer further determines how strongly each emitting region contributes to the image.

Cunningham examined the effects of redshift and focusing on the spectrum of a disk around a Kerr black hole~\cite{Cunningham1975}, while Luminet constructed the image of a spherical black hole illuminated by a thin disk~\cite{Luminet1979}. Later studies clarified the distinction between direct emission, lensing rings, and photon rings through the winding of null trajectories and their intersections with the emitting material~\cite{GrallaHolzWald2019}. Close to the capture zone, the logarithmic growth of the deflection angle produces a hierarchy of increasingly narrow images~\cite{Bozza2002}. The interferometric signatures and shape of these structures have motivated tests of the spacetime geometry through the photon ring~\cite{Johnson2020,GrallaLupsascaMarrone2020,Paugnat2022}. Their relative brightness, however, also depends on the extent and optical properties of the emitting flow~\cite{Vincent2022}.

Thin disk calculations have examined boson and Proca stars~\cite{RosaRubiera2022}, regular black holes and geonic horizonless configurations~\cite{OlmoRosa2023}, and alternative spherical geometries through their photon ring contributions~\cite{DiasDaSilva2023}. Analyses of nonsingular de Sitter core geometries and black holes surrounded by dark matter have further shown how the emission prescription affects the apparent differences between spacetimes~\cite{DeMartino2023,MacedoRosa2024}. More recently, images of parametrized black holes have extended these comparisons to families of exterior metrics using several Standard Unbound emission profiles~\cite{OlmoRosa2025}.

Non--commutative geometry introduces a fundamental scale into the localization of spacetime events. Snyder's quantized spacetime provided an early formulation in terms of non--commuting coordinates~\cite{Snyder1947}, while spacetime uncertainty relations connected localization with quantum mechanics and gravity~\cite{Doplicher1995}. The resulting field theories depend on the coordinate algebra, product, and symmetry prescription, as reviewed by Douglas and Nekrasov and by Szabo~\cite{DouglasNekrasov2001,Szabo2003}.

Several prescriptions implement non--commutativity in gravity. In gauge formulations based on the star product, the Seiberg Witten map relates deformed gauge fields to their commutative counterparts~\cite{SeibergWitten1999}. Gravitational constructions apply this procedure to the tetrad and spin connection~\cite{Chamseddine2001Deforming,Chaichian2008Schwarzschild}. Twisted diffeomorphisms provide a related formulation of differential geometry and gravitational actions~\cite{Aschieri2005Gravity,Aschieri2006Geometry}. In source smearing, Einstein's geometric field equations keep their classical form, and an extended matter distribution carries the non--commutative scale. Coordinate coherent states motivate Gaussian localization~\cite{SmailagicSpallucci2003Coherent,Nicolini2006}, while Lorentzian distributions provide a prescription with an algebraic radial tail~\cite{Anacleto2020}.

Within the Gaussian prescription, Nicolini, Smailagic, and Spallucci obtained a Schwarzschild inspired geometry with a regular central region and a modified evaporation endpoint~\cite{Nicolini2006}. This construction was extended to charged configurations with smeared mass and charge distributions~\cite{Ansoldi2007Charged} and to rotating geometries~\cite{SmailagicSpallucci2010Kerrr}. Scalar Hawking emission was also examined for higher dimensional non--commutative inspired black holes~\cite{NicoliniWinstanley2011}. The broader relation between source smearing, black hole geometry, and thermodynamics was reviewed in Ref.~\cite{Nicolini2009Review}. Gaussian and Lorentzian profiles have different radial tails: their corrections to the asymptotic Schwarzschild lapse decay exponentially and algebraically, respectively~\cite{Nicolini2006,Anacleto2020}.

For a Lorentzian matter distribution, Anacleto et al. examined the absorption and scattering of massless scalar waves and identified changes in the corresponding cross sections~\cite{Anacleto2020}. Their investigations also addressed corrections to Schwarzschild and BTZ black hole entropy in the presence of the generalized uncertainty principle~\cite{Anacleto2021SchwarzschildEntropy,Anacleto2021BTZEntropy}. Campos et al. analyzed scalar quasinormal modes and the reduction of the shadow radius~\cite{Campos2022}, while the extension to a global monopole background connected absorption, scattering, and shadow observables within the same Lorentzian prescription~\cite{Anacleto2023Monopole}. More recently, non--commutative thin shell gravastars have provided a horizonless setting for the analysis of photon trajectories and deflection~\cite{Anacleto2026Gravastar}.

Related investigations by Ara\'ujo Filho et al. examined thermodynamics, perturbations, and weak and strong gravitational lensing in the Lorentzian non--commutative geometry~\cite{AraujoFilho2024Properties}, together with horizons, shadows, and scalar perturbations in its axisymmetric extension~\cite{AraujoFilho2025Axisymmetric}. Geodesics, accretion emission, retrolensing, time delay, and neutrino propagation were considered in Refs.~\cite{AraujoFilho2025Accretion,AraujoFilho:2025rzh}. Furthermore, the study of periodic massive particle trajectories connected the displacement of characteristic orbital scales with gravitational wave signatures and an estimate from the S2 orbit~\cite{Heidari2026}.

Juri\'c, Kumara, and Po\v{z}ar constructed non--commutative black holes with several Moyal twists and examined their different effects on horizons and curvature~\cite{Juric2025Constructing}. Touati proposed another construction in which the Seiberg Witten map acts on interaction potentials before the gravitational field equations are solved~\cite{Touati2026Potentials}. Within gauge based deformations, Heidari, Ara\'ujo Filho, and Lobo studied Hayward spacetime~\cite{Heidari2025Hayward}, while subsequent investigations considered Kalb Ramond and bumblebee black holes~\cite{AraujoFilho2025KalbRamond,AraujoFilho2026BumblebeeNC}. Photon trajectories, shadows, and weak and strong lensing were further analyzed in the non--commutative Kalb Ramond background~\cite{AraujoFilho2026OpticalKalbRamond}.

Particle production in a Lorentzian non--commutative spacetime has been studied through Bogoliubov transformations and tunneling, together with evaporation and greybody factors~\cite{AraujoFilho2025ParticleProduction}. In addition, Ara\'ujo Filho and Lobo clarified the role of the horizon location and the surface gravity prescription in tunneling and thermodynamic calculations for axially symmetric geometries~\cite{AraujoFilhoLobo2025TunnelingComment,AraujoFilhoLobo2026ThermodynamicsComment}. More recently, Ara\'ujo Filho et al. examined bosonic and fermionic particle creation, emission rates, and evaporation lifetimes for a Schwarzschild like geometry constructed with a temporal angular Moyal twist and well defined surface gravity~\cite{AraujoFilho2026QuantumFeatures}.

In this work, we examine a transparent, geometrically thin emitting disk in the effective exterior geometry supplied by a Lorentzian mass distribution. We compare black holes at fixed asymptotic mass using the same radial emission parameters and normalization. From the photon trajectories and transfer functions, we calculate the displacement, peak brightness, and integrated contribution of the image orders for static and neutral circular emitters. In addition, we vary the emission profile and include the inner boundary of stable circular motion. These calculations connect the narrow lensed features with the critical curve and the broad direct contribution with the emitting region. We also compare the characteristic scales with those of the unexpanded Lorentzian lapse and describe the Reissner Nordstr\"om correspondence of the effective metric.


\section{Optical scales of the non--commutative geometry}
\label{sec:nc-optical}

Initially, we determine the horizon, photon orbit, critical impact parameter, and innermost stable circular orbit of the effective non--commutative geometry. These scales set the capture threshold and the inner boundary of stable circular emission used in the disk images.

We work in geometrized units, $G=c=1$, and use the static, spherically symmetric line element
\begin{equation}
\mathrm{d}s^2=-f(r)\,\mathrm{d}t^2+\frac{\mathrm{d}r^2}{f(r)}+r^2\left(\mathrm{d}\vartheta^2+\sin^2\vartheta\,\mathrm{d}\varphi^2\right).
\label{eq:nc-line-element}
\end{equation}

The source is smeared according to a Lorentzian distribution. This prescription implements non--commutativity through the matter source~\cite{Nicolini2006,Anacleto2020}, with the associated mass function given by~\cite{Heidari2026}
\begin{equation}
\mathcal{M}(r)=\frac{2M}{\pi}\left[\arctan\!\left(\frac{r}{\sqrt{\pi\Theta}}\right)-\frac{r\sqrt{\pi\Theta}}{r^2+\pi\Theta}\right],\qquad f_{\mathrm{L}}(r)=1-\frac{2\mathcal{M}(r)}{r}.
\label{eq:nc-exact-lapse}
\end{equation}
Here, $M$ is the asymptotic mass and $\Theta\geq0$ has dimensions of length squared. The subscript $\mathrm{L}$ identifies the unexpanded Lorentzian lapse. Following the orbital analysis of Ref.~\cite{Heidari2026}, we introduce $x=r/M$, $\theta=\Theta/M^2$, and $\zeta=8\sqrt{\theta/\pi}$. The leading exterior expansion gives
\begin{equation}
f(x)=1-\frac{2}{x}+\frac{\zeta}{x^2},\qquad f_{\mathrm{L}}(x)=f(x)-\frac{16\sqrt{\pi}\,\theta^{3/2}}{3x^4}+\mathcal{O}\!\left(\frac{\theta^{5/2}}{x^6}\right).
\label{eq:nc-effective-lapse}
\end{equation}

The expansion parameter is $\sqrt{\pi\theta}/x$. Equation~\eqref{eq:nc-effective-lapse} describes the exterior where this ratio is small, while Eq.~\eqref{eq:nc-exact-lapse} also describes the regular central geometry. We construct the images with $f$ and use $f_{\mathrm{L}}$ to compare the characteristic exterior scales.

The effective geometry possesses an outer horizon at $x_+=1+\sqrt{1-\zeta}$, with $0\leq\zeta\leq1$, or equivalently $0\leq\theta\leq\pi/64$. This interval specifies the horizon domain of the effective metric. For photons, the circular orbit condition is $x f'(x)-2f(x)=0$, where a prime denotes differentiation with respect to $x$. The unstable exterior photon orbit and critical impact parameter are
\begin{equation}
x_{\mathrm{ph}}=\frac{3+\sqrt{9-8\zeta}}{2},\qquad \beta_{\mathrm{c}}=\frac{b_{\mathrm{c}}}{M}=\frac{x_{\mathrm{ph}}}{\sqrt{f(x_{\mathrm{ph}})}}.
\label{eq:nc-critical}
\end{equation}

The photon orbit specifies a radius in the spacetime, and $\beta_{\mathrm{c}}$ gives the critical curve radius in impact coordinates. On the other hand, the visible edge of a brightness depression depends on both photon propagation and the radial distribution of emission~\cite{GrallaHolzWald2019,PerlickTsupko2022}.

For neutral circular emitters, the innermost stable circular orbit is determined by the exterior root
\begin{equation}
x_{\mathrm{ISCO}}^3-6x_{\mathrm{ISCO}}^2+9\zeta x_{\mathrm{ISCO}}-4\zeta^2=0,\qquad x_{\mathrm{ISCO}}>x_{\mathrm{ph}}.
\label{eq:nc-isco}
\end{equation}

\begin{figure}[tbp]
\centering
\includegraphics[width=0.5\linewidth]{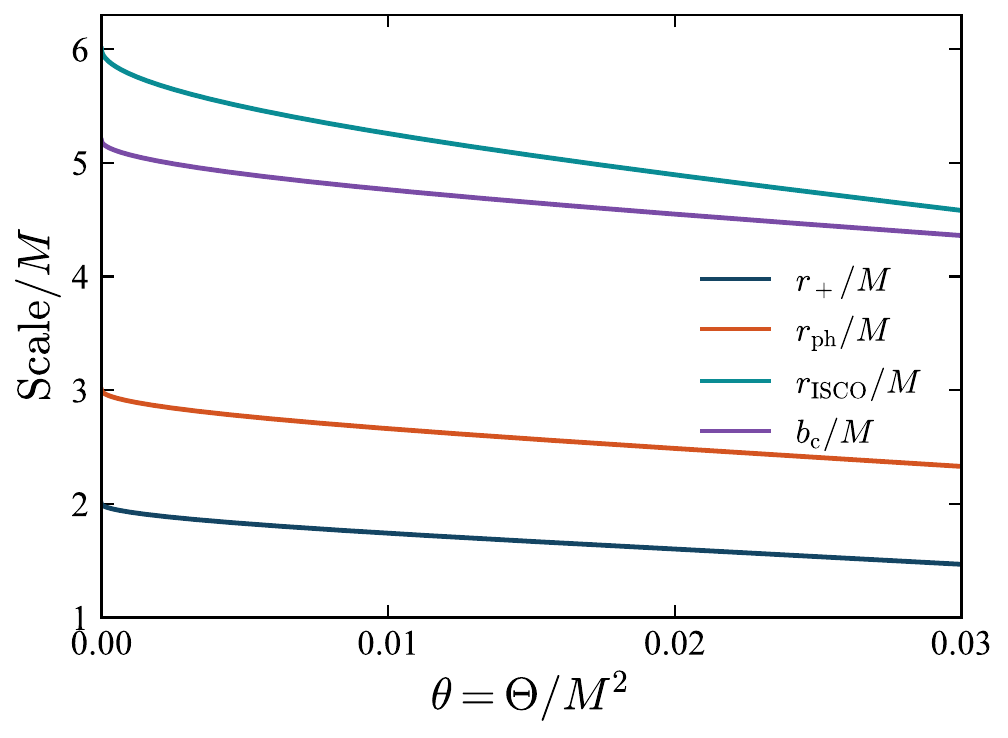}
\caption{Characteristic optical and orbital scales of the effective non--commutative geometry as functions of $\theta=\Theta/M^2$. All lengths are divided by the fixed asymptotic mass $M$. The photon orbit radius $r_{\mathrm{ph}}$ and the critical impact parameter $b_{\mathrm{c}}$ characterize the orbital and screen scales, respectively.}
\label{fig:nc-scales}
\end{figure}

Figure~\ref{fig:nc-scales} shows that the horizon, photon orbit, ISCO, and critical impact parameter decrease throughout the parameter interval considered. Table~\ref{tab:nc-scales} gives their numerical values. At $\theta=0.03$, the critical curve is approximately $16.1\%$ smaller than in Schwarzschild at fixed $M$, and the ISCO radius decreases by $23.7\%$. This contraction agrees with the trend obtained from null geodesics in Ref.~\cite{Campos2022}. In this manner, increasing non--commutativity reduces the photon capture scale and allows stable circular emitters to extend closer to the black hole.

\begin{table}[tbp]
\centering
\caption{Characteristic scales of the effective metric. The final column compares its critical impact parameter with that obtained from the unexpanded Lorentzian lapse, $\Delta b_{\mathrm{c}}/b_{\mathrm{c,L}}=(b_{\mathrm{c}}-b_{\mathrm{c,L}})/b_{\mathrm{c,L}}$.}
\label{tab:nc-scales}
\begin{tabular}{rrrrrr}
\toprule
$\theta$ & $r_+/M$ & $r_{\mathrm{ph}}/M$ & $r_{\mathrm{ISCO}}/M$ & $b_{\mathrm{c}}/M$ & $100\Delta b_{\mathrm{c}}/b_{\mathrm{c,L}}$ \\
\midrule
0.00 & 2.00000 & 3.00000 & 6.00000 & 5.19615 & $0.000$ \\
0.01 & 1.74071 & 2.66073 & 5.25673 & 4.76285 & $-0.030$ \\
0.02 & 1.60141 & 2.48660 & 4.89426 & 4.54807 & $-0.115$ \\
0.03 & 1.46716 & 2.32854 & 4.58045 & 4.35966 & $-0.283$ \\
\bottomrule
\end{tabular}
\end{table}

The unexpanded Lorentzian lapse gives nearby exterior scales over the interval studied. At $\theta=0.03$, the expansion parameter at the effective photon orbit is approximately $0.132$. Replacing $f$ by $f_{\mathrm{L}}$ increases the photon orbit radius by approximately $0.67\%$ and the critical impact parameter by $0.28\%$; the outer horizon difference is approximately $1.50\%$, with the unexpanded values used as denominators. The image calculations below resolve how rays near the critical curve intersect the emitting disk.


\section{Photon trajectories and disk transfer functions}
\label{sec:nc-rays}

Spherical symmetry allows each photon trajectory to be described in a plane through the center. Let $E=-k_t$ and $L$ be its conserved energy and total angular momentum, and define $b=L/E$ and $\beta=b/M$. The null condition gives
\begin{equation}
\left(\frac{\mathrm{d}r}{\mathrm{d}\lambda}\right)^2=E^2\left[1-\frac{b^2f(r)}{r^2}\right],\qquad \frac{\mathrm{d}\chi}{\mathrm{d}\lambda}=\frac{L}{r^2},
\label{eq:nc-null-first}
\end{equation}
where $\lambda$ is an affine parameter and $\chi$ measures the angle within the photon plane. For numerical integration, it is convenient to set $u=M/r$ and use the equivalent equations
\begin{equation}
\left(\frac{\mathrm{d}u}{\mathrm{d}\chi}\right)^2=\frac{1}{\beta^2}-u^2+2u^3-\zeta u^4,\qquad \frac{\mathrm{d}^2u}{\mathrm{d}\chi^2}=-u+3u^2-2\zeta u^3.
\label{eq:nc-null-u}
\end{equation}

The equation of second order passes smoothly through radial turning points and follows the incoming and outgoing portions of the trajectory in one integration. Backward integration ends at the outer horizon, at escape to infinity, or at the prescribed angular limit. The dimensionless residual of the first relation shows a direct numerical check. For the radial trajectories used here it remains below $1.2\times10^{-12}$. Independently, trajectories computed at high precision agree with the radial scattering integral to better than $10^{-24}$.

\begin{figure}[tbp]
\centering
\includegraphics[width=\linewidth]{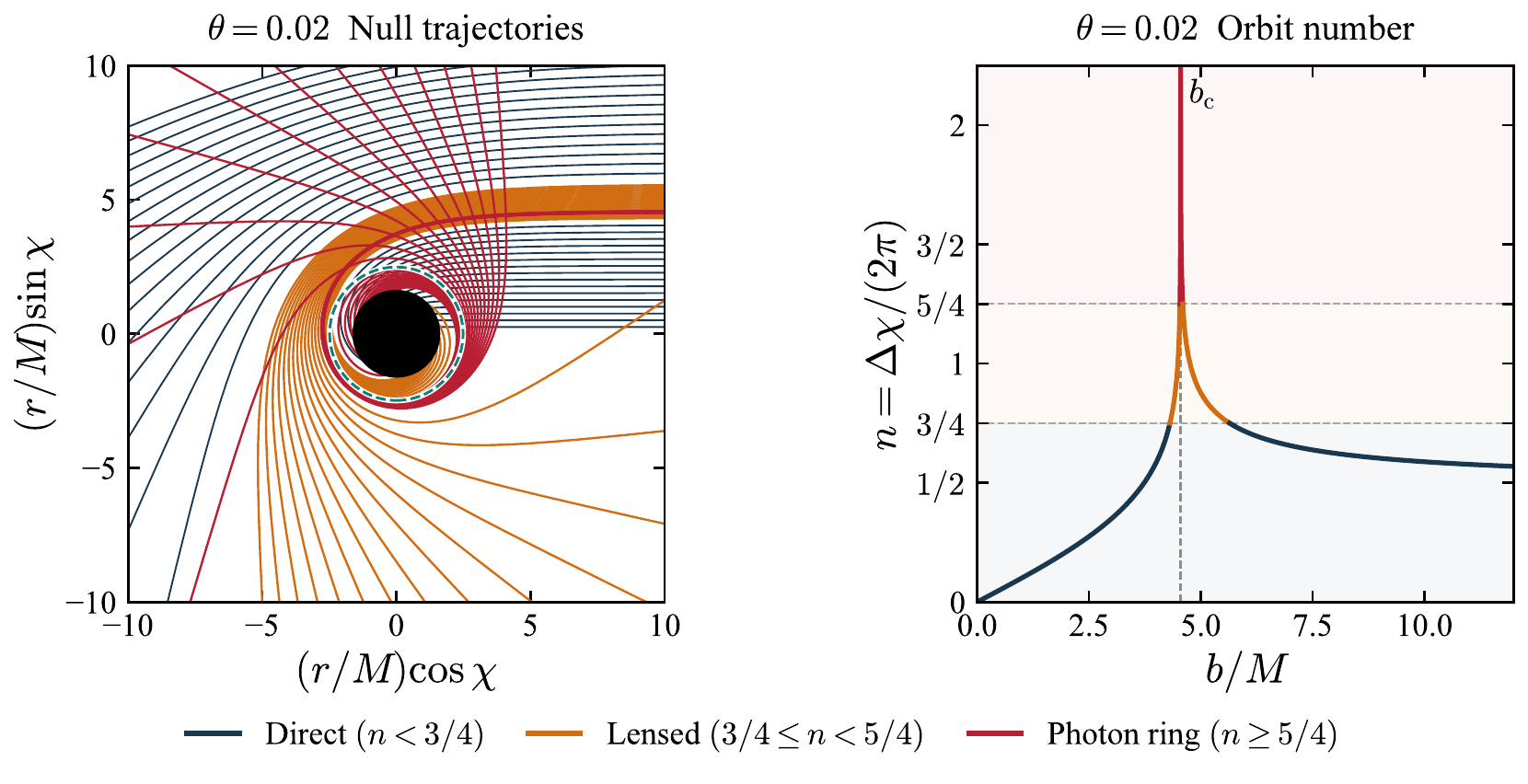}
\caption{$82$ photon trajectories and the orbit number for $\theta=0.02$, with the same winding colors in both panels: direct rays in blue, lensed rays in orange, and the photon ring class in red. Incoming photons approach from the right in the trajectory plane. The horizon is the filled black disk, and the dashed circle marks the photon orbit. Capture and scattering are separated by $\beta_{\mathrm{c}}=4.54807$; both outcomes occur within each winding class.}
\label{fig:nc-ray-families}
\label{fig:nc-rays}
\label{fig:nc-orbit-number}
\end{figure}

We describe the angular advance through the orbit number $n_{\mathrm{orb}}=\Delta\chi/(2\pi)$~\cite{GrallaHolzWald2019}. Figure~\ref{fig:nc-ray-families} displays the trajectories and orbit number for $\theta=0.02$. In the trajectory plane, $\mathcal{X}=x\cos\chi$ and $\mathcal{Y}=x\sin\chi$; incident photons approach from positive $\mathcal{X}$ along the asymptotes $\mathcal{Y}=\beta$. Reflection across the horizontal axis gives the opposite angular orientation. The black disk marks the horizon at $x_+=1.60141$, and the dashed circle marks the unstable photon orbit at $x_{\mathrm{ph}}=2.48660$.

Photons with $\beta<\beta_{\mathrm{c}}=4.54807$ are captured and terminate at the horizon. Those with $\beta>\beta_{\mathrm{c}}$ reach a radial turning point and are scattered back to infinity. The colors additionally describe the winding of each trajectory: direct trajectories satisfy $n_{\mathrm{orb}}<3/4$, lensed trajectories satisfy $3/4\leq n_{\mathrm{orb}}<5/4$, and the photon ring class satisfies $n_{\mathrm{orb}}\geq5/4$~\cite{GrallaHolzWald2019}. Each winding class contains captured and scattered trajectories. In particular, a photon can circle close to the photon orbit before entering the horizon or returning to large radius. As $\beta$ approaches $\beta_{\mathrm{c}}$ from either side, the angular accumulation diverges, producing the behavior in the right panel. For the displayed geometry, the photon ring class extends approximately over $4.53229\leq\beta\leq4.59654$, with capture and scattering on the respective sides of $\beta=\beta_{\mathrm{c}}$. At the critical value, the limiting trajectory approaches the unstable circular orbit asymptotically.

The orbit number measures the complete trajectory from infinity to the horizon for capture, or from infinity back to infinity for scattering. The intersection index counts successive encounters with the emitting sheet. A ray with large winding can collect radiation at several intersections, with each contribution weighted by the local emissivity and frequency shift.

The observer is static at $x_{\mathrm{o}}=1000$. We write $\mathsf{X}=X/M$ and $\mathsf{Y}=Y/M$ for the screen impact coordinates, so that $\beta^2=\mathsf{X}^2+\mathsf{Y}^2$. The local angular separation $\alpha_{\mathrm{sky}}$ from the inward radial direction obeys $\sin\alpha_{\mathrm{sky}}=\beta\sqrt{f(x_{\mathrm{o}})}/x_{\mathrm{o}}$. The inclination $i$ is measured from the disk normal. In the screen convention used here, intersections with the equatorial emitting sheet satisfy
\begin{equation}
\cos i\cos\chi+\frac{\mathsf{Y}}{\beta}\sin i\sin\chi=0,\qquad \chi_m=\chi_1+(m-1)\pi,\qquad m=1,2,\ldots.
\label{eq:nc-crossings}
\end{equation}

The first positive solution defines $\chi_1$. For an axial observer, $i=0$, this reduces to $\chi_m=(m-1/2)\pi$. We consider inclinations $0\leq i<\pi/2$, for which intersections with the infinitesimally thin emitting sheet are geometrically well defined. Throughout this analysis, $m=1$ denotes the direct intersection, $m=2$ the first higher order intersection, and larger $m$ successive contributions along the same backward ray.

\begin{figure}[tbp]
\centering
\includegraphics[width=0.5\linewidth]{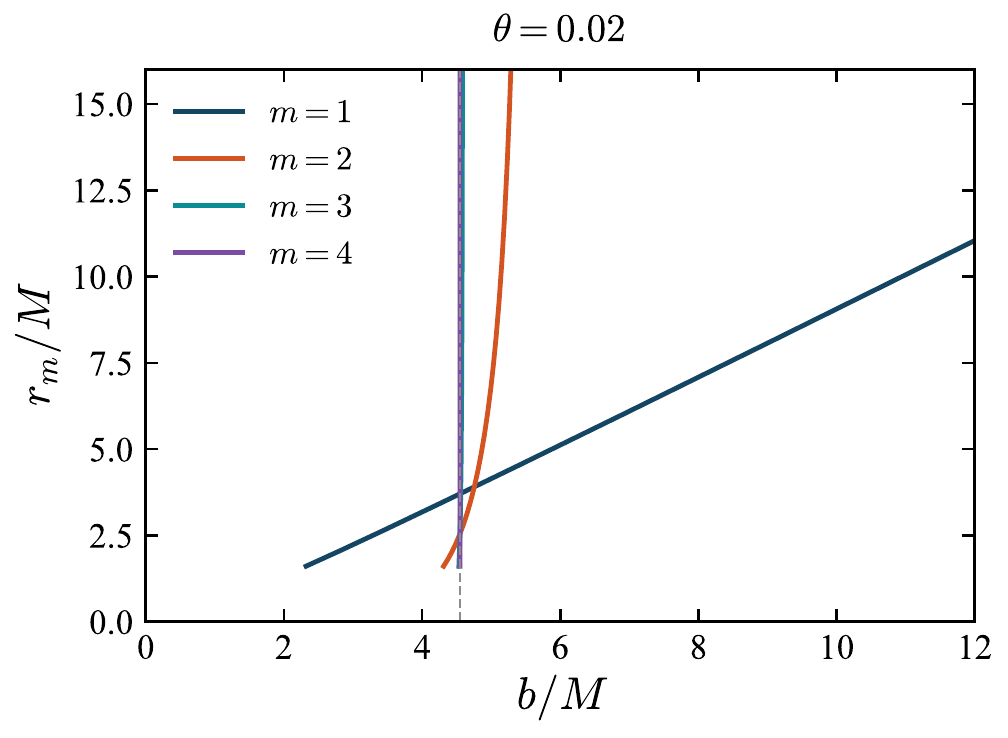}
\caption{Axial transfer functions $x_m=r_m/M$ for the non--commutative geometry with $\theta=0.02$. Each function gives the emission radius at the corresponding intersection with the disk plane as a function of impact parameter. Its domain comprises the rays that reach that intersection. The intersection index begins at $m=1$ for the direct contribution.}
\label{fig:nc-transfer}
\end{figure}

The transfer function is the radius at which the $m$th intersection occurs,
\begin{equation}
x_m(\beta)=\frac{1}{u\bigl[(m-1/2)\pi;\beta\bigr]},\qquad i=0,
\label{eq:nc-transfer}
\end{equation}
with its domain supplied by rays that reach the corresponding angle before termination. Figure~\ref{fig:nc-transfer} shows these functions for $\theta=0.02$. The direct transfer function maps a broad interval of emission radii onto the screen, while the higher order functions concentrate around the critical impact parameter. Their steep slopes map a finite radial interval of the disk into a narrow image interval. In this manner, the transfer functions explain the hierarchy of lensed features~\cite{GrallaHolzWald2019,GrallaLupsascaMarrone2020,OlmoRosa2023}.

For photons scattered from infinity back to infinity, we define the deflection angle as $\widehat{\alpha}=\Delta\chi-\pi$. This asymptotic configuration gives an independent check on the integration, complementing the finite observer position used for the images. Expanding the null geodesic deflection integral at large impact parameter gives
\begin{equation}
\widehat{\alpha}(\beta)=\frac{4}{\beta}+\frac{3\pi(5-\zeta)}{4\beta^2}+\mathcal{O}(\beta^{-3}).
\label{eq:nc-weak}
\end{equation}

\begin{figure}[tbp]
\centering
\includegraphics[width=\linewidth]{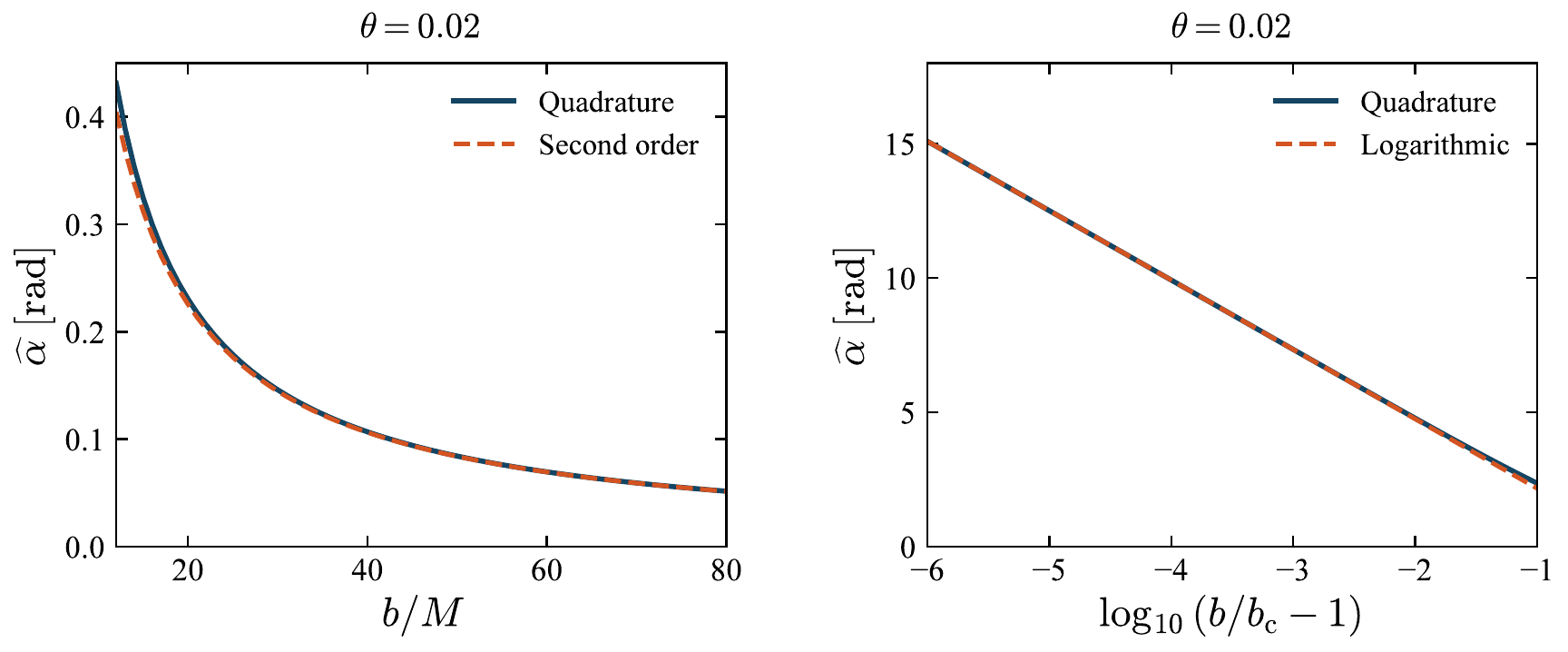}
\caption{Numerical deflection and asymptotic lensing checks for the non--commutative geometry at $\theta=0.02$. The left panel tests the large impact parameter expansion, while the right panel tests the logarithmic behavior close to the critical curve. Both use infinity to infinity scattering.}
\label{fig:nc-lensing}
\end{figure}

The non--commutative correction reduces the deflection at second order for fixed $\beta$. Close to the critical curve, the deflection develops the logarithmic behavior~\cite{Bozza2002}
\begin{equation}
\widehat{\alpha}(\beta)=-\overline{a}\log\!\left(\frac{\beta}{\beta_{\mathrm{c}}}-1\right)+\overline{b}+\mathcal{O}\!\left[\left(\frac{\beta}{\beta_{\mathrm{c}}}-1\right)\log\!\left(\frac{\beta}{\beta_{\mathrm{c}}}-1\right)\right],\qquad \overline{a}=\frac{x_{\mathrm{ph}}}{\sqrt{x_{\mathrm{ph}}^2-2\zeta}}.
\label{eq:nc-strong}
\end{equation}
Here, $\overline{b}$ is the finite part after subtraction of the logarithmic divergence. The resulting hierarchy is the spherical counterpart of the increasingly narrow subrings discussed in Ref.~\cite{Johnson2020}. The coefficient $\overline{a}$ increases from unity in Schwarzschild to approximately $1.1854$ at $\theta=0.03$. Successive disk image separations from the critical curve scale as $\exp(-\pi/\overline{a})$ per additional half orbit. It is worth pointing out that the increase in $\overline{a}$ gives a larger ratio between successive separations, even as the critical curve moves inward. Figure~\ref{fig:nc-lensing} compares the numerical deflection with the expansion at large impact parameter and the logarithmic expression near the critical curve.

\section{Emission prescriptions and observed intensity}
\label{sec:nc-emission}

We model the disk as a transparent, geometrically thin emitting sheet with prescribed isotropic bolometric surface intensity in the emitter frame. Each disk intersection contributes to the observed radiation, and photons propagate freely between intersections. The emission and velocity profiles determine the source contribution, while the metric determines the trajectories and frequency shifts~\cite{GrallaLupsascaMarrone2020,OlmoRosa2023,DiasDaSilva2023,MacedoRosa2024}.

We employ the Johnson SU profile used in phenomenological disk imaging~\cite{GrallaLupsascaMarrone2020},
\begin{equation}
\mathcal{J}(x;\gamma,\mu,\sigma)=\frac{\exp\!\left[-\frac{1}{2}\left(\gamma+\operatorname{arcsinh}\frac{x-\mu}{\sigma}\right)^2\right]}{\sqrt{(x-\mu)^2+\sigma^2}},\qquad I_{\mathrm{em}}(x)=\frac{I_0}{\mathcal{N}}\mathcal{J}(x;\gamma,\mu,\sigma)\,\mathbf{1}_{[x_{\mathrm{in}},40]}(x).
\label{eq:nc-emissivity}
\end{equation}

The dimensionless parameters $\mu$ and $\sigma>0$ locate the profile and control its radial width, while $\gamma$ determines its asymmetry. The indicator function fixes the emitting interval, and $\mathcal{N}$ sets the normalization. For the families tied to the horizon, $\mathcal{N}$ is the maximum of $\mathcal{J}$ over that interval, so $I_0$ is the peak emitted intensity. We find this maximum by setting $y=\operatorname{arcsinh}[(x-\mu)/\sigma]$ and solving $\gamma+y+\tanh y=0$, with the stationary point restricted to the chosen support.

\begin{figure}[tbp]
\centering
\includegraphics[width=0.5\linewidth]{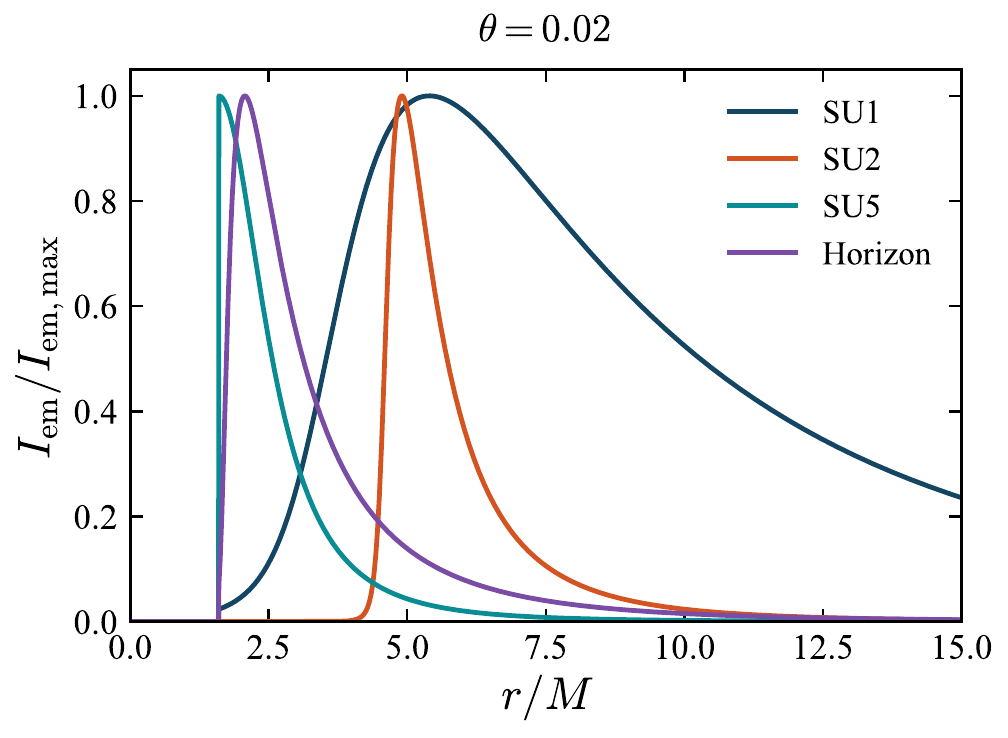}
\caption{Representative normalized emission profiles for $\theta=0.02$. These families tied to the horizon illustrate the dependence on the location and width of the source. The comparison between different geometries in Figs.~\ref{fig:nc-intensity} and \ref{fig:nc-axial} instead uses a single fixed SU2 profile, with $(\gamma,\mu,\sigma)=(-2,17/3,1/4)$ and common normalization.}
\label{fig:nc-emission}
\end{figure}

Figure~\ref{fig:nc-emission} compares representative emission prescriptions at $\theta=0.02$. The numerical model includes ten SU families of the type considered in Ref.~\cite{DeMartino2023}, together with profiles whose inner boundary lies at the horizon, photon orbit, or ISCO. The families tied to $x_+$ follow the horizon displacement as $\theta$ changes. For the main comparison between geometries, we fix $(\gamma,\mu,\sigma)=(-2,17/3,1/4)$ and use a common normalization evaluated on $2\leq x\leq40$. This fixed SU2 profile peaks outside the ISCO for all four geometries. In this regard, the main emitting region keeps the same radial placement, and the horizon cutoff acts on its innermost tail.

The transfer of intensity follows from the conservation of $I_\nu/\nu^3$ along a vacuum null geodesic~\cite{Cunningham1975,Vincent2022}. If $g=\nu_{\mathrm{o}}/\nu_{\mathrm{e}}$, the specific intensity and the intensity integrated over frequency transform according to
\begin{equation}
I_{\nu,\mathrm{o}}(\nu_{\mathrm{o}})=g^3 I_{\nu,\mathrm{e}}(\nu_{\mathrm{o}}/g),\qquad I_{\mathrm{o}}=\int I_{\nu,\mathrm{o}}\,\mathrm{d}\nu_{\mathrm{o}}=g^4 I_{\mathrm{em}}.
\label{eq:nc-liouville}
\end{equation}
The fourth power of $g$ follows from integrating the specific intensity over the observed frequency. All intensity plots use this bolometric expression.

For supported static emitters, the redshift factor and the total observed intensity are
\begin{equation}
g_{\mathrm{stat}}(x)=\sqrt{\frac{f(x)}{f(x_{\mathrm{o}})}},\qquad I_{\mathrm{obs}}(\mathsf{X},\mathsf{Y})=\sum_{m=1}^{m_{\max}}g_m^4 I_{\mathrm{em}}(x_m).
\label{eq:nc-static-intensity}
\end{equation}
Every intersection contributes once because the sheet is transparent. For an axial observer, each term is $f(x_m)^2 I_{\mathrm{em}}(x_m)/f(x_{\mathrm{o}})^2$. The factor $f(x_m)^2$ attenuates emission near the horizon. In this manner, static emitters isolate the gravitational contribution to the frequency shift.

To examine the effect of orbital motion, we also consider neutral emitters on equatorial circular geodesics. Their angular velocity and temporal component of the four velocity are
\begin{equation}
M\Omega(x)=\frac{\sqrt{x-\zeta}}{x^2},\qquad u^t(x)=\left(1-\frac{3}{x}+\frac{2\zeta}{x^2}\right)^{-1/2}.
\label{eq:nc-circular-motion}
\end{equation}
The source occupies $x\geq x_{\mathrm{ISCO}}$, corresponding to the region of stable circular motion. With $\ell_z=k_\varphi/(ME)=-\mathsf{X}\sin i$, the frequency shift becomes
\begin{equation}
g_{\mathrm{circ}}(x,\mathsf{X},i)=\frac{1}{\sqrt{f(x_{\mathrm{o}})}\,u^t(x)\left[1-M\Omega(x)\ell_z\right]}.
\label{eq:nc-circular-redshift}
\end{equation}
This expression includes gravitational redshift, transverse time dilation, and the Doppler shift along the line of sight. At $i=0$, orbital motion still enters through $u^t$. We use the same fixed SU2 parameters and normalization for static and circular emitters, preserving the common amplitude after the ISCO cutoff. The resulting comparison describes the combined effects of orbital motion and the inner boundary of stable circular emission.


\section{Disk images and dependence on non--commutativity}
\label{sec:nc-images}

\begin{figure}[tbp]
\centering
\includegraphics[width=0.5\linewidth]{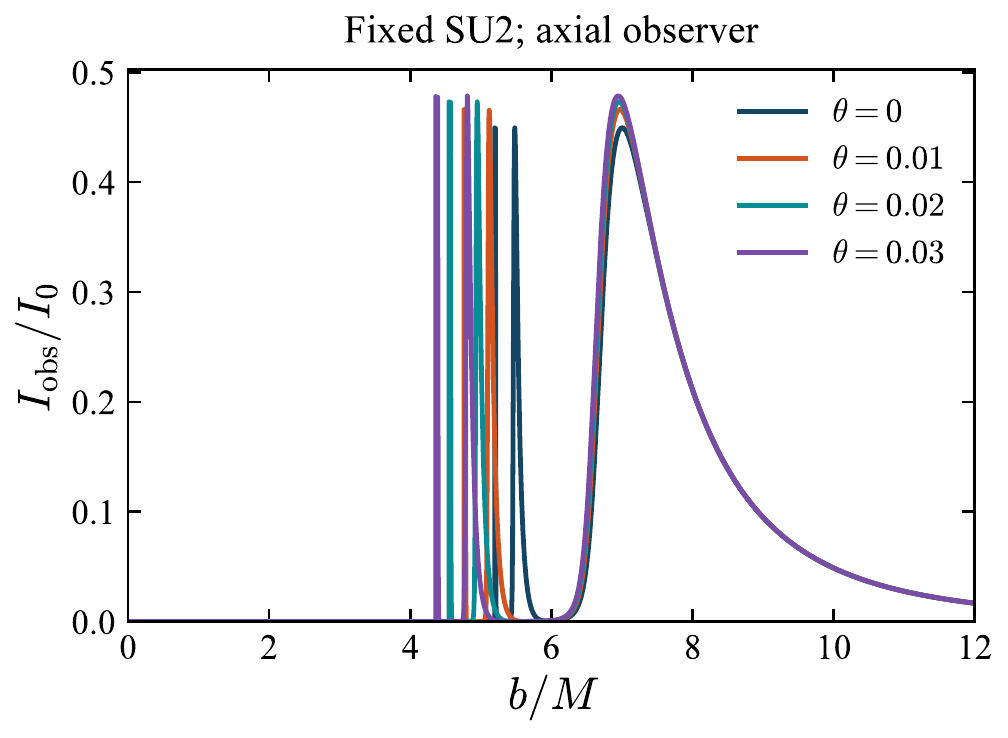}
\caption{Axial bolometric intensity for the same fixed SU2 source in four geometries. Intensities use the common emitted scale $I_0$. The critical impact parameters change with $\theta$, while the radial parameters of the dominant emitting region are held fixed. The plotted curves use a uniform impact spacing of $0.008$ supplemented by logarithmic samples on both sides of $\beta_{\mathrm{c}}$.}
\label{fig:nc-intensity}
\end{figure}

Figure~\ref{fig:nc-intensity} presents the axial intensity profiles for the fixed SU2 source at $\theta=0,0.01,0.02,0.03$. The geometries share the asymptotic mass, emission parameters, outer radius, and intensity normalization. The narrow higher order features move inward with the critical curve, while the broad direct contribution follows the location and width of the emitting region. In this manner, the profiles distinguish the response of the lensed image scales from that of the dominant direct emission.

The numerically refined maxima quantify the distinct responses of the direct and lensed images. As shown in Table~\ref{tab:nc-image-measures}, the direct peak moves from $\beta\simeq7.004$ to $6.947$ between the Schwarzschild case and the $\theta=0.03$ geometry, a displacement of less than $1\%$. Moreover, the first higher order peak moves from $\beta\simeq5.481$ to $4.809$, which follows more closely the contraction of the critical curve. The peak direct intensity increases by approximately $6.5\%$ for this source. To compare the overall radial signal within a fixed aperture, we additionally compute
\begin{equation}
\mathcal{S}(\beta_{\max})=2\pi\int_0^{\beta_{\max}}\frac{I_{\mathrm{obs}}(\beta)}{I_0}\,\beta\,\mathrm{d}\beta,\qquad \beta_{\max}\simeq18.
\label{eq:nc-screen-signal}
\end{equation}
This quantity is the intensity integrated over the screen in impact coordinates. In the asymptotic limit of small angles, the angularly integrated flux is proportional to $I_0(M/r_{\mathrm{o}})^2\mathcal{S}$. Within the adopted aperture, $\mathcal{S}$ increases by approximately $4.3\%$ between $\theta=0$ and $0.03$ for the fixed static SU2 source. At $\theta=0.02$, direct emission supplies approximately $96.51\%$ of the signal, the first higher order term supplies $3.34\%$, and the remaining orders contribute $0.15\%$. It is important to mention that the appreciable peak brightness of a narrow lensed feature is accompanied by a small integrated contribution because it occupies a narrow interval on the screen.

\begin{table}[tbp]
\centering
\caption{Axial image measures for the fixed static SU2 source. The peak positions are refined by solving the corresponding transfer relation; $I_{1,\max}$ gives the maximum of the direct contribution, while $\mathcal{S}$ includes up to seven disk crossings over the same aperture, $\beta_{\max}\simeq18$.}
\label{tab:nc-image-measures}
\begin{tabular}{rrrrr}
\toprule
$\theta$ & $\beta_{1,\max}$ & $\beta_{2,\max}$ & $I_{1,\max}/I_0$ & $\mathcal{S}$ \\
\midrule
0.00 & 7.004 & 5.481 & 0.4496 & 42.37 \\
0.01 & 6.971 & 5.119 & 0.4662 & 43.42 \\
0.02 & 6.957 & 4.949 & 0.4732 & 43.86 \\
0.03 & 6.947 & 4.809 & 0.4786 & 44.21 \\
\bottomrule
\end{tabular}
\end{table}

The radial calculations include up to seven intersections. Increasing the sum from five to seven changes $\mathcal{S}$ by less than $5.6\times10^{-7}$ in relative terms across the four geometries. At $\theta=0.02$, refining the impact grid from $5852$ to $11701$ points at the same aperture changes $\mathcal{S}$ by $8.0\times10^{-6}$.

\begin{figure}[tbp]
\centering
\includegraphics[width=\linewidth]{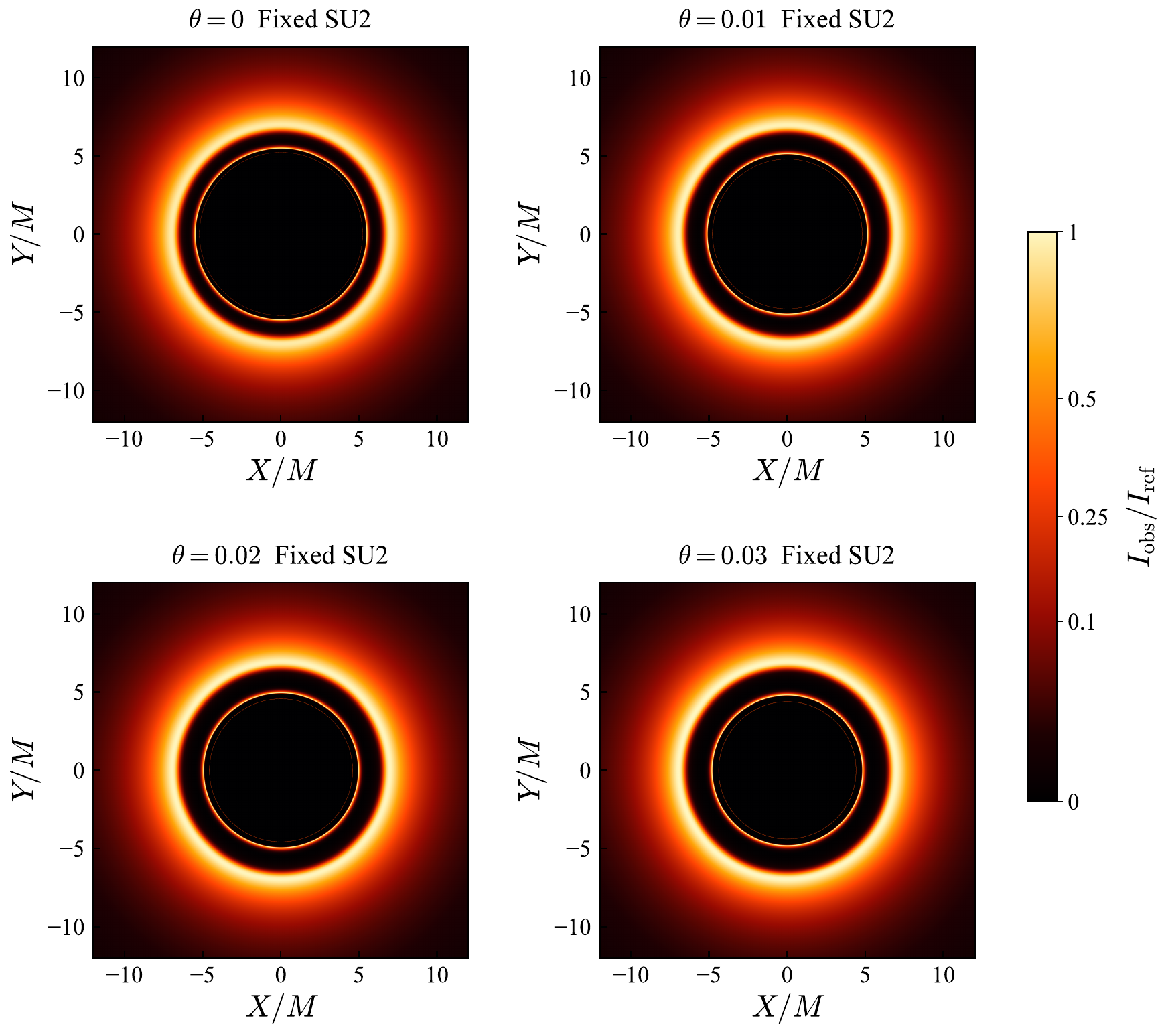}
\caption{Axial images of the fixed SU2 disk for $\theta=0,0.01,0.02,0.03$, ordered from the upper left to the lower right. The panels share the emission prescription, asymptotic mass, field of view, and brightness normalization.  The plotted screen coordinates are impact parameters divided by $M$.}
\label{fig:nc-axial}
\end{figure}

Figure~\ref{fig:nc-axial} shows the axial images with a common field of view and brightness scale. We display $\sqrt{I_{\mathrm{obs}}/I_{\mathrm{ref}}}$, with $I_{\mathrm{ref}}=0.478644I_0$, while the intensity curves use the linear intensity. The common reference preserves the relative brightness changes between geometries. The maps contain $2401\times2401$ pixels with a $2\times2$ subpixel average and separate radial interpolation on each side of $\beta_{\mathrm{c}}$. The narrow lensed structures contract with the critical curve. In addition, the faint emission assigned to the inner disk by the fixed SU2 profile determines the extent of the central depression.

The brightness changes follow from the frequency shift and the transfer functions. At fixed emission radius, $\partial f/\partial\zeta=x^{-2}>0$, increasing the static redshift factor $g^4=f^2/f(x_{\mathrm{o}})^2$ in the emitting region. Moreover, varying $\zeta$ changes the radii $x_m$ sampled at a fixed screen position. The local slope of the emission profile determines whether this radial redistribution enhances or suppresses the corresponding contribution.

\begin{figure}[tbp]
\centering
\includegraphics[width=\linewidth]{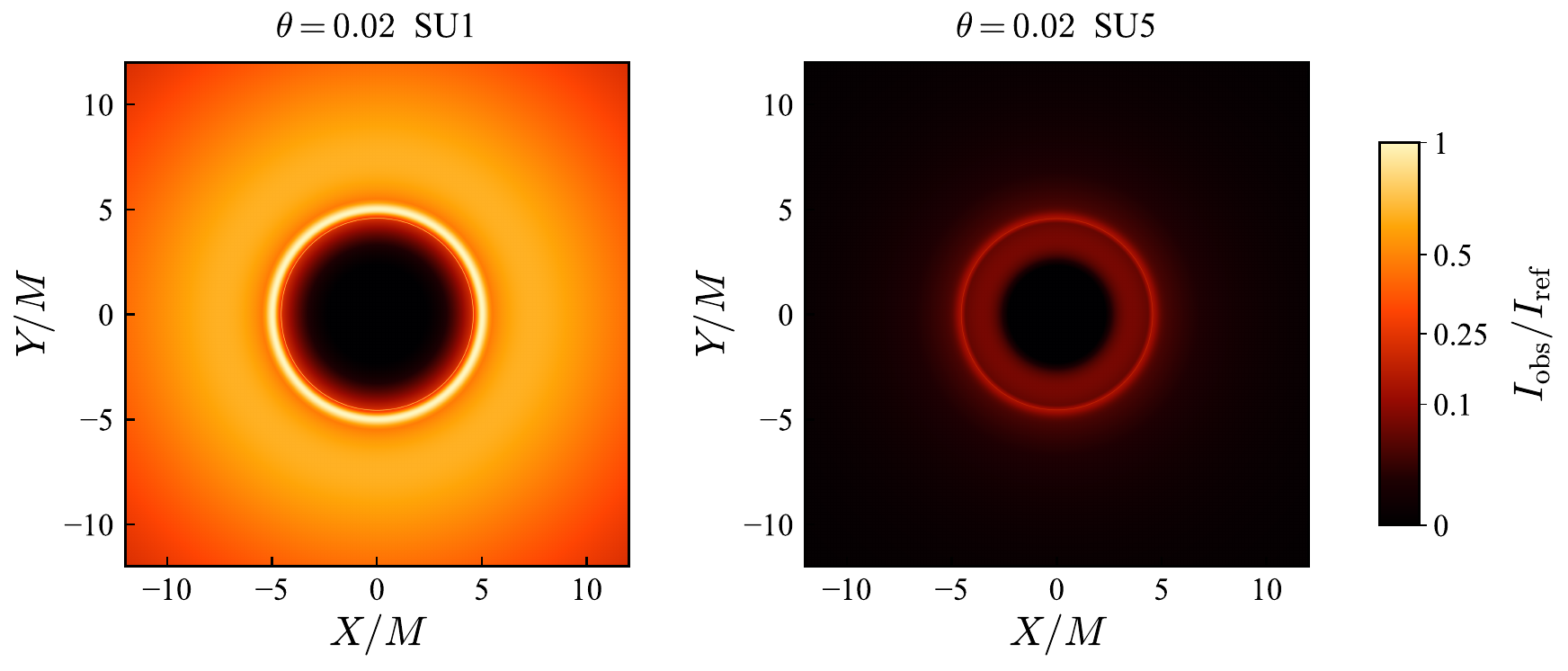}
\caption{Axial images for static emitters with the SU1 (left) and SU5 (right) prescriptions tied to the horizon at the same $\theta=0.02$. Each local profile is normalized to the same peak emitted scale $I_0$. The panels share the geometry and critical curve, and the common exterior color scale uses the reference $I_{\mathrm{ref}}=0.702671I_0$. Their different brightness patterns arise from the radial distribution of emission.}
\label{fig:nc-source-images}
\end{figure}

Figure~\ref{fig:nc-source-images} compares the SU1 and SU5 profiles tied to the horizon at fixed $\theta=0.02$. The broader SU1 distribution and the more concentrated SU5 profile weight the same transfer functions differently. Emission near the horizon fills regions that remain faint for the outer fixed SU2 source, with gravitational attenuation set by $f^2$. Both profiles have the same peak emitted intensity, while their different radial extents determine the total emitted power. Their common critical curve and distinct brightness patterns show directly how the source distribution shapes the central depression.

\begin{figure}[tbp]
\centering
\includegraphics[width=\linewidth]{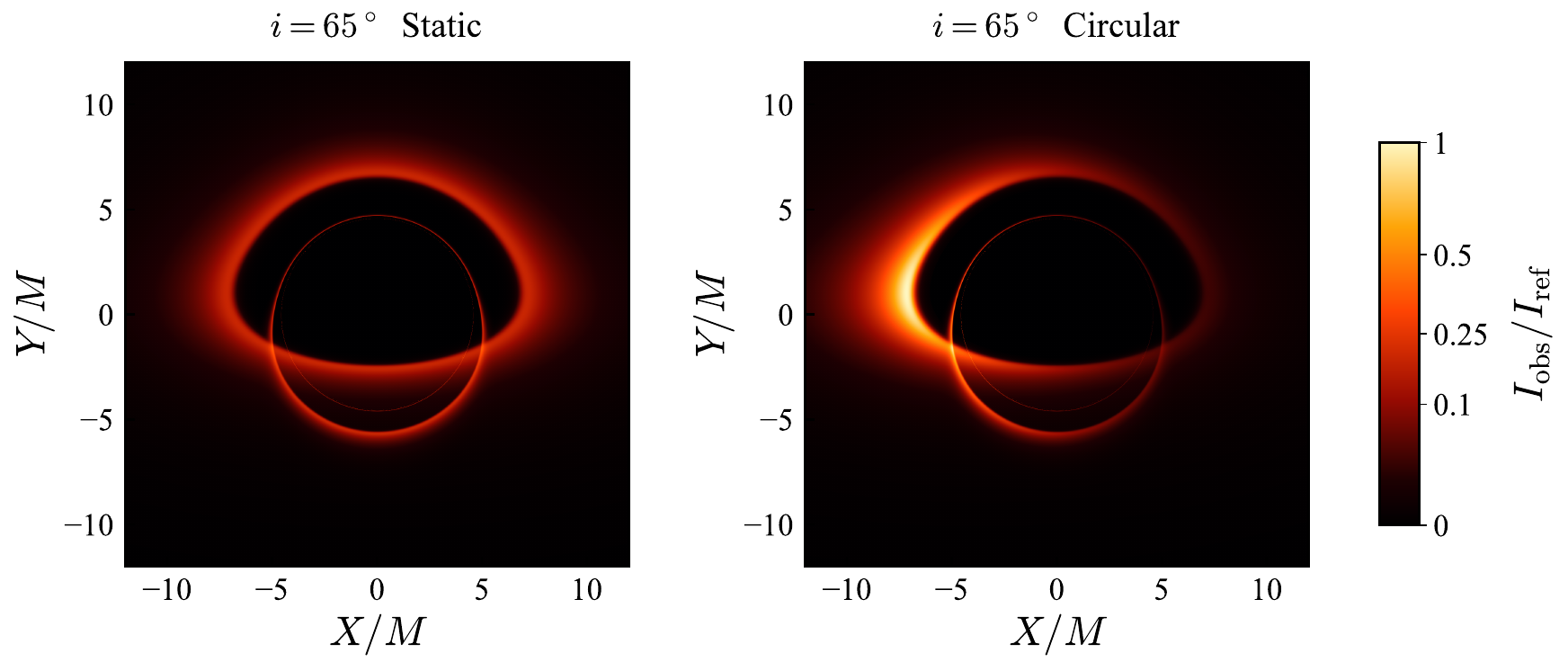}
\caption{Inclined images at $\theta=0.02$ and $i=65^\circ$ for supported static emitters (left) and neutral circular emitters in the region $r\geq r_{\mathrm{ISCO}}$ (right). Both use the fixed SU2 profile and the same emitted and displayed intensity scales. The image for circular emitters takes into account orbital time dilation and Doppler modulation in addition to gravitational redshift. The $1501\times1501$ maps evaluate the null trajectories at each pixel center, and the common exterior color scale uses $I_{\mathrm{ref}}=2.47720I_0$.}
\label{fig:nc-inclined}
\end{figure}

Figure~\ref{fig:nc-inclined} compares static and circular emitters at $\theta=0.02$ and $i=65^\circ$. Static emission preserves symmetry between the left and right sides. For circular emitters, Doppler modulation brightens the approaching side relative to the receding side. Furthermore, transverse time dilation changes the observed intensity, and the ISCO sets the inner boundary of stable circular emission.

\FloatBarrier


\section{Conclusion}
\label{sec:conclusion}

We examined the optical appearance of a non--commutative Schwarzschild black hole surrounded by a transparent thin emitting disk. From the effective exterior metric supplied by a Lorentzian mass distribution, we calculated photon trajectories, disk transfer functions, and bolometric intensities for static and neutral circular emitters.

Increasing $\theta$ moved the horizon, photon orbit, and innermost stable circular orbit inward. Between $\theta=0$ and $0.03$, the critical curve contracted by approximately $16.1\%$ at fixed asymptotic mass, and the ISCO radius decreased by $23.7\%$. The narrow higher order features followed this contraction. On the other hand, the broad direct peak of the fixed SU2 source moved by less than $1\%$, while the first higher order peak shifted from $b/M\simeq5.481$ to $4.809$.

With the same static SU2 profile and common normalization, the peak direct intensity increased by approximately $6.5\%$, and the signal integrated over the adopted aperture increased by $4.3\%$. At $\theta=0.02$, direct emission supplied approximately $96.51\%$ of the integrated signal. The narrow lensed features combined appreciable local brightness with a small integrated contribution. Moreover, changing the radial emissivity produced distinct central depressions at the same critical curve. For inclined circular emitters, Doppler modulation generated the asymmetry between approaching and receding sides, while time dilation and the ISCO cutoff also shaped the images.

The effective lapse coincided with the Reissner Nordstr\"om form under $Q_{\mathrm{eff}}^2/M^2=\zeta$, giving identical trajectories and images for identical neutral emission prescriptions. At $\theta=0.03$, the unexpanded Lorentzian lapse changed the critical impact parameter by approximately $0.28\%$. This comparison supported the exterior scale trends found in the effective geometry. Further applications of this spacetime include the Magnus effect and spin precession~\cite{Nishida:2026bzu,Geng:2026xcs}.


\section*{Acknowledgments}
\hspace{0.5cm} A. A. Araújo Filho is supported by Conselho Nacional de Desenvolvimento Cient\'{\i}fico e Tecnol\'{o}gico (CNPq) -- [150223/2025-0]. This work by A.~F.~Santos is partially supported by National Council for Scientific and Technological Development - CNPq project No. 312406/2023-1.
V. B. Bezerra is partially supported by the National Council for Scientific and Technological Development (CNPq) - Brazil, Project number 311847/2026-9.

\section*{Data Availability Statement}

The analytical expressions and numerical figures are presented in the manuscript. The code used to generate the numerical results and figures is available from the corresponding author upon reasonable request.

\bibliography{main}

\end{document}